\documentclass[12pt,a4paper]{article}
\usepackage{amssymb}
\usepackage{amsmath}
\usepackage{amsfonts,latexsym}

\newcommand{\nohat}{}
\newcommand{\noth}{}
\newcommand{\x}{\times}

\begin{document}

\rightline{HU-EP-01/51}
\rightline{UG-01/35}
\rightline{UUITP-01/07}
\vskip 1truecm

\centerline{\LARGE \bf Massive Dualities in Six Dimensions}
\vskip 1truecm

\centerline{\bf K.~Behrndt}
\centerline{Institut f\"ur Physik, Humboldt University}
\centerline{Invalidenstra\ss e 110, 10115 Berlin, Germany}
\centerline{E-mail: {\tt behrndt@physik.hu-berlin.de}}
\vskip .5truecm

\centerline{\bf E.~Bergshoeff and D.~Roest}
\centerline{Institute for Theoretical Physics, University of Groningen}
\centerline{Nijenborgh 4, 9747 AG Groningen, The Netherlands}
\centerline{E-mail: {\tt e.a.bergshoeff@phys.rug.nl, d.roest@phys.rug.nl}}
\vskip .5truecm

\centerline{\bf P.~Sundell}
\centerline{Department of Theoretical Physics, Uppsala University}
\centerline{Box 803, S-75108 Uppsala, Sweden}
\centerline{E-mail: {\tt Per.Sundell@teorfys.uu.se}}
\vskip 2truecm

\centerline{ABSTRACT}
\bigskip

We study compactifications of string theory and M-theory to six dimensions
with background fluxes. The nonzero fluxes lead to additional mass
parameters. We derive the S-- and T--duality rules for the
corresponding (massive) supergravity theories.
Specifically, we investigate the massive
T-duality between Type IIA superstring theory compactified
on K3 with background fluxes and Type IIB superstring theory
compactified on K3. Furthermore,
we generalise to the massive case the 6D 'string-string' S-duality
between
M-theory on K3 $\times S^1$
and the Heterotic String on $T^4$.
Whereas in the case of massive T--duality the mass
parameters are in the fundamental representation of the U--duality group
$O(4,20)$ we find that in the case of massive S--duality they
are in the 3-index antisymmetric representation. In the latter case
the mass parameters involved extend those of 
Kaloper and Myers. We apply our duality rules to massive
brane solutions, like the  domain wall solutions corresponding to
the mass parameters and find new massive brane solutions.
Finally, we discuss the higher-dimensional interpretation
of the dualities and brane solutions.

\vfill\eject

\section{Introduction}

It is well-known that M-theory and the five string theories are related
via different T-- and S--dualities.
In the low energy limit this manifests itself by relations between
the corresponding supergravities or $S^1$-compactifications thereof.
When considering compactifications to six dimensions other dualities
arise. For instance, Type IIA and Type IIB string theory compactified on
K3 and the Heterotic string theory compactified on $T^4$ are related
by various dualities \cite{Hull:1995ys, Witten:1995ex,Duff:1994zt}. 
At the level of the
corresponding D=6, N=2 supergravities the explicit form of the different
duality relations were derived in \cite{Behrndt:1996si}. The relations are
illustrated in Figure 1.
\vskip .8truecm

\begin{figure}[h]

\begin{picture}(300,210)(-75,-10)

\put(100,200){\makebox(0,0){\large \bf M}}
\put(0,150){\makebox(0,0){\large \bf H}}
\put(200,150){\makebox(0,0){\large \bf IIA}}
\put(300,150){\makebox(0,0){\large \bf IIB}}
\put(100,50){\makebox(0,0){\large \bf m}}
\put(0,0){\makebox(0,0){\large \bf h}}
\put(200,0){\makebox(0,0){\large \bf iia}}
\put(300,0){\makebox(0,0){\large \bf iib}}

\put(90,190){\line(-2,-1){75}}\put(50,190){\makebox(0,0){\large $S^1/\mathbb{Z}_2$}}
\put(110,190){\line(2,-1){75}}\put(150,190){\makebox(0,0){\large $S^1$}}
\put(110,40){\line(2,-1){75}}\put(150,40){\makebox(0,0){\large $S^1$}}

\put(0,140){\line(0,-1){130}}\put(-15,75){\makebox(0,0){\large $T^4$}}
\put(200,140){\line(0,-1){130}}\put(185,75){\makebox(0,0){\large $K3$}}
\put(300,140){\line(0,-1){130}}\put(285,75){\makebox(0,0){\large $K3$}}
\put(100,190){\line(0,-1){130}}\put(85,125){\makebox(0,0){\large $K3$}}

\put(100,0){\vector(-1,0){70}\vector(1,0){140}\makebox(-140,-17){\large ${\cal S}$}}
\put(250,0){\vector(-1,0){30}\vector(1,0){60}\makebox(-60,-17){\large ${\cal T}$}}
\put(250,150){\vector(-1,0){30}\vector(1,0){60}\makebox(-60,-17){\large ${\cal T}$}}
\put(50,50){\vector(-1,0){30}\vector(1,0){60}\makebox(-60,-17){\large ${\cal S}$}}

\end{picture}

\it \caption{Web of dualities. The $S^1$ reduction of M-theory
to the Type IIA superstring may be seen
as an S-duality since the strong coupling limit of the Type IIA superstring
corresponds to a large radius limit
of the $S^1$. The $S^1/\mathbb{Z}_2$ reduction of M-theory to the
Heterotic model is the Horava-Witten
scenario \cite{Horava:1995qa}. In the lower line
{\bf h}, {\bf iia} and {\bf iib} are six-dimensional theories
whereas {\bf m} is seven-dimensional. The
higher-dimensional origin is indicated by vertical lines.
The duality relations between the theories are
indicated by horizontal two-sided arrows with
$\cal S$ denoting an S--duality and
${\cal T}$ indicating a T--duality.}

\end{figure}
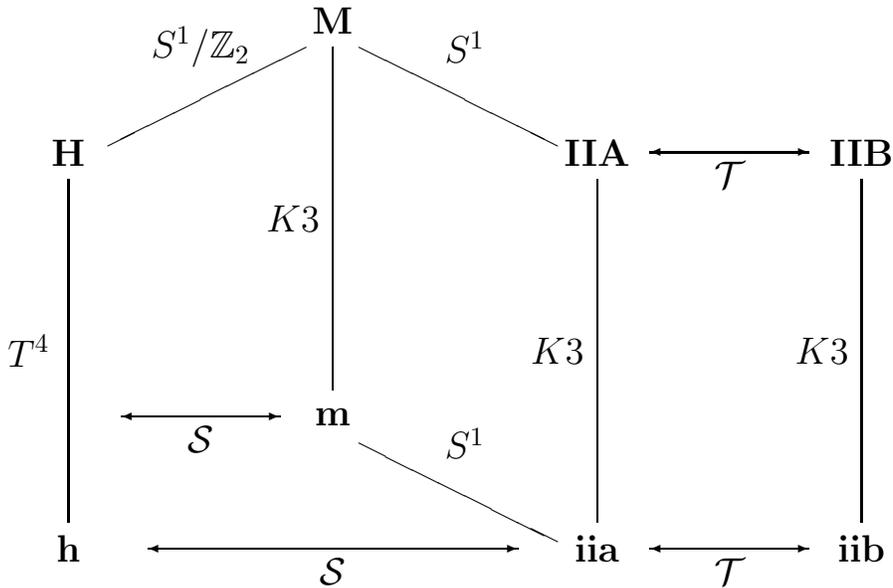

In this work we consider an extension of the results of
\cite{Behrndt:1996si} where we apply, instead of ordinary
Kaluza-Klein reductions, Scherk-Schwarz reductions or, more generally,
compac\-ti\-fi\-cations with non-zero background fluxes. This allows us
to extend the six-dimensional dualities
to duality relations involving a set of mass parameters $\{m\}$.
The new rules constitute
a massive deformation of the old rules in the sense that for $\{m=0\}$ we
recover the duality rules of \cite{Behrndt:1996si}.

Compactifications with background fluxes have attracted a renewed
interest in view of the fact that they lead to massive supergravities
which play an important role in (i) the AdS/CFT correspondence
\cite{Maldacena:1998re}, (ii) the Randall-Sundrum scenario
\cite{Randall:1999ee,Randall:1999vf} and (iii) recent
cosmological applications, see e.g.~\cite{Townsend:2001ea, Kallosh:2001gr}.
These types of compactifications are a generalization of
Scherk-Schwarz (SS) reductions \cite{Scherk:1979zr} where a non-trivial
background flux is given to some field strength $F_{p+1}$
of a p-form tensor field $A_p$. The nonzero values are taken in
the directions of a non-trivial (p+1)-cycle of the compactification
manifold.

The case $p=0$ where $A_0$ is the R-R axionic scalar of IIB supergravity
has been considered in \cite{Bergshoeff:1996ui} and generalised in
\cite{Lavrinenko:1998qa, Meessen:1999qm}. Similarly, the NS-NS axionic
scalars of the Heterotic model were used in \cite{Kaloper:1999yr} to generate
massive deformations of its toroidal reductions. In all these cases the same
deformations can be obtained directly in the lower-dimensional theory by
the gauging of a subgroup of the U-duality group. The mass parameters are
the structure constants of the gauged subgroup in some basis
\cite{Bergshoeff:1997mg, Alonso-Alberca:2000gh}.
More recently, the SS-reduction of
massive IIA supergravity \cite{Romans:1986tz}
on K3 with background fluxes has been considered
\cite{Haack:2001iz}. The resulting massive supergravity theory
in six dimensions
does not seem to correspond to the gauging of any global symmetry.

In generalized SS reductions the following subtlety plays an
important role.
Consider the reduction on a product manifold $A \times B$.
For an ordinary Kaluza-Klein (KK) reduction the order of
compactification on
A and B is of no influence. First compactify on $A$ and next on
$B$ or the
other way around gives the same (massless) supergravities. It is
 said that
two KK reductions commute with each other.
The situation changes for SS reductions \cite{Cowdall:1997tw}.
Here the order does matter in general, i.e.~reducing over an internal
manifold $A \times B$ gives a different result than
when reducing over $B \times A$. The different ways of performing
SS reductions lead to supergravities with different massive deformations.
Therefore {\it SS reductions do not commute} in general.

The dependence on the reduction scheme arises since one
throws away Fourier modes on the internal manifolds.
In a KK reduction all but the zero-mode
is thrown away, in which case the order of compactification
is irrelevant.
On the other hand, in a SS reduction one keeps a combination of the
zero-mode and higher Fourier modes, depending on the mass parameter.
In such a  case the order in which one
throws away the higher-order
Fourier modes is relevant and leads to different results in general.

A natural question is whether the massive deformations of the
supergravities
preserve the duality relations. The T-duality between Type IIA and Type
IIB string theory in D=10 has
been generalised to the massive case involving
a single mass parameter \cite{Bergshoeff:1996ui}. However,
the eleven-dimensional origin of this mass parameter is not well
understood. Due to this one cannot construct a massive
S-duality between massive IIA and D=11 supergravity.
Recently, the T-duality between massive IIA supergravity compactified
on K3 \cite{Haack:2001iz} and IIB supergravity compactified
on K3 (see Figure 1) has been extended to the massive
case, as derived by \cite{Janssen:2001hy} and, independently,
by the present authors.

It has been found in the SS
reduction of Type IIA string theory on K3 \cite{Haack:2001iz}
 that the mass parameters fill
multiplets of the U-duality group, which is perturbative in this case.
Similarly, it has been found that in SS reductions of Type IIA
string theory on $T^2$ and $T^4$ \cite{Singh:2001gt},
the mass parameters fill multiplets of the perturbative part of the
U-duality group.
In this sense the non-perturbative part of the U-duality group is broken,
while the perturbative part survives as a so-called pseudo-symmetry,
i.e.~the Lagrangian is invariant provided we also transform the
mass parameters.

The prevailing theme seems to be
 that perturbative dualities can be extended
to the massive case but that
for non-perturbative dualities this is problematic.
In this work we will consider massive extensions of the duality
relations in
six dimensions. Indeed the T-duality between Type IIA and Type IIB
string theory compactified on K3 can be generalised to
include mass parameters, while the S-duality between Type IIA string
theory compactified on K3 and the Heterotic model
on $T^4$ is more subtle, as argued in
\cite{Haack:2001iz,Janssen:2001hy}. Remarkably,
we are able to construct a massive S-duality in six dimensions by
considering
SS reductions in a different order. An intriguing feature is that
the massive S-duality contains a set of mass parameters that
are in the 3-index antisymmetric representation
of the U-duality group $O(4,20)$ whereas the
massive T-duality involves a set of mass parameters that
are in the fundamental representation of $O(4,20)$.
It remains an
open issue whether there are massive dualities for
{\it all} massive deformations occuring in compactifications
of string theory and/or M-theory.

This work is organised as follows. In Section 2 we derive the massive
T-duality rules between massive Type IIA and Type IIB string theory
on K3.
The same duality rules have, independently, been
derived in \cite{Janssen:2001hy}. Furthermore, we apply the massive
T-duality
rules to massive brane solutions. Next, in Section 3 we construct
massive S-duality rules between M-theory compactified on K3$\times S^1$
and the Heterotic string theory compactified on $T^4$.
Again we apply these massive S-duality rules to brane solutions.
In Section 4 we consider the higher-dimensional interpretation of the 1/2
BPS brane solutions in D=6. In particular, we discuss the domain
wall solutions and discuss their relations
under duality. Finally, we will conclude with a Discussion and Summary
Section. Our notational conventions are given in Appendix
A.  In Appendix B we will consider the SS reduction of the Heterotic string
theory on $T^4$. This result will be needed in order to construct the
massive S-duality rules in
Section 3 and generalises the work of \cite{Kaloper:1999yr}.

\section{Massive T-duality}

In this Section we consider the extension to the massive case
of the T-duality in six dimensions
beween Type IIA and Type IIB string theory compactified on K3.
The contents of this section overlaps with \cite{Janssen:2001hy}
where a similar analysis was performed.

\subsection{Supergravity Relations}

The Type IIA theory, compactified on K3 to six dimensions,
 has the following massless field content:
\begin{align}
  \text{IIA:} \qquad \big\{ \nohat{g}, \nohat{B}, \nohat{\phi},
  \nohat{V}^A, \nohat{M}^{AB} \big\} \,.
\label{fciia}
\end{align}
Here $g$ is the metric, $B$ is the NS-NS 2-form and $\phi$ is the dilaton.
The $A,B$ indices ($A,B = 1, \cdots 24$) refer to the U-duality group
$O(4,20)$. The 1-forms $V^A$ are in the fundamental representation of
$O(4,20)$. The scalar matrix $\nohat{M}^{AB}$ parametrises the coset
$O(4,20)/O(4) \times O(20)$. It contains  80 = 4 (dilatonic) + 76
(axionic) scalars. Being an element of $O(4,20)$ it satisfies
\begin{equation}
M^{-1}_{AD} = L_{AB} M^{BC} L_{CD}\, .
\end{equation}
The $O(4,20)$ metric $L_{AB}$ is iteratively defined by
\begin{equation}
  L_{AB} = L^{AB} = \left( \begin{array}{ccc}
  0 & 1 & 0\\
  1 & 0 & 0\\
  0 & 0 & L_{ab} \end{array} \right)\, ,
\end{equation}
where $L_{ab}$ ($a,b = 1, \cdots 22$) is the $O(3,19)$ metric. Similarly,
the scalar matrix $M^{AB}$ is iteratively defined by
\begin{equation}
  M^{AB} =
  \left(
  \begin{array}{ccc}
  -e^{-2\sigma} + \ell_a \ell_b { { M}}^{ab}
  -\tfrac{1}{4} e^{2\sigma} {\ell}^4&
  -\tfrac{1}{2} e^{2{\sigma}}{\ell}^2&
  {\ell}_a{ M}^{ab}  -\tfrac{1}{2} e^{2{\sigma}}{\ell}^2
  {\ell}^b \cr
  -\tfrac{1}{2} e^{2{\sigma}}{\ell}^2&
  -e^{2\sigma}&
  e^{2{\sigma}}{\ell}^b\cr
  { M}^{ab}{\ell}_b - \tfrac{1}{2} e^{2{\sigma}}{\ell}^2
  {\ell}^a &
  e^{2{\sigma}}{\ell}^a&
  { M}^{ab} - e^{2\sigma}{\ell}^a{\ell}^b
  \end{array}
  \right)\, .
\end{equation}
while $M^{kl} = L^{kl} = (\mathbb{I}_{16})^{kl}$ with $k,l=1..16$.
Here the  $\ell_a$ are 22 axionic scalars ($\ell^2 = \ell_a \ell^a$),
$\sigma$ is a dilatonic scalar and $M^{ab}$ parametrizes the coset
$O(3,19)/O(3)\times O(19)$.
The $O(4,20)$-duality symmetry acts as
\begin{align}
  V^A \rightarrow \Omega^A_{\; B} V^B \,, \qquad
  M^{AB} \rightarrow \Omega^A_{\; C} \Omega^B_{\; D} M^{CD} \,, \qquad
\label{O(4,20)}
\end{align}
with $\Omega^A_{\; B} \in O(4,20)$.
This group consists of $276$ generators, of which
$76$ shift the axions with a constant: $\ell \rightarrow \ell + m$.

In a recent paper the Scherk-Schwarz generalized reduction of
ten-dimensional massive IIA supergravity \cite{Romans:1986tz}
on K3 was performed \cite{Haack:2001iz}. This compactification yields
the Lagrangian
\begin{align}
  \mathcal{L}_{\text{IIA,6}} = & \tfrac{1}{2} \sqrt{-\nohat{g}}
e^{-2\nohat{\phi}}
    \Big[ -\nohat{R} + 4 |d \nohat{\phi}|^2 -\tfrac{3}{4}
|d \nohat{B}|^2
    +\tfrac{1}{8} \text{Tr}(d \nohat{M} d \nohat{M}^{-1})
    -e^{2\nohat{\phi}} |\nohat{F}^A|^{2}
    - 2 e^{2\nohat{\phi}} |m^A|^{2} \Big]  \notag \\[2mm]
  & \qquad -\tfrac{1}{8} \nohat{\epsilon}
    \Big[ \nohat{B} \nohat{F}^A \nohat{F}_A
    - m^A \nohat{B}^2 \nohat{F}_A +
    \tfrac{1}{3} m^2 \nohat{B}^3 \Big] \,,
\label{actionmiia}
\end{align}
with the field strength
\begin{align}
  \nohat{F}^A \equiv d \nohat{V}^A + m^A \nohat{B} \,.
\label{mfs}
\end{align}
We use a notation where the $O(4,20)$-indices in $|F^A|^2$ and $|m^A|^2$
are contracted using the scalar matrix $M^{-1}$, e.g.
\begin{equation}
|m^A|^2  = m^A (M^{-1})_{AB}\, m^B\, .
\end{equation}

The Scherk-Schwarz reduction introduces a total of 24 mass parameters $m^A$.
Note that the NS-NS two-form $\nohat{B}$ becomes massive in the
presence of fluxes.
Furthermore, the nonzero $m^A$ break the $O(4,20)$ symmetry
to a pseudo-symmetry \cite{Haack:2001iz}:
\begin{align}
  V^A \rightarrow \Omega^A_{\; B} V^B \,, \qquad
  M^{AB} \rightarrow \Omega^A_{\; C} \Omega^B_{\; D} M^{CD} \,, \qquad
  m^A \rightarrow \Omega^A_{\; B} m^B \,,
\end{align}
with $\Omega^A_{\; B} \in O(4,20)$. The 24 mass parameters
fill the fundamental multiplet $m^A$.

We wish to compare the 6D IIA Lagrangian with the 6D IIB Lagrangian
that results from the KK compactification of IIB supergravity on K3.
In order to do so, it is convenient to bring the 6D IIB Lagrangian
into a more convenient form.
In the fluxless case the six-dimensional IIB
action can be written in a manifest $O(5,21)$ invariant form.
The model contains 5 self-dual and 21 anti-selfdual two-forms with
corresponding duality relations.
To match the IIA field content it is convenient to extract one
self-dual and one anti-self-dual two-form from the total of 26.
This gives us the unconstrained NS-NS two-form $\nohat{B}$ plus
4 selfdual and 20 anti-selfdual 2-forms.
The 105 scalars of the theory are  organized into a dilaton $\nohat{\phi}$,
24 axionic scalars
$\nohat{\ell}^A$ while the remaining 80 scalars parametrise
a $O(4,20)/O(4)\times O(20)$ coset matrix $\nohat{M}^{AB}$.
Thus the field content consists of
\begin{align}
  \text{IIB:} \qquad \big\{ \nohat{g}, \nohat{B}, \nohat{\phi},
  \nohat{B}^A, \nohat{\ell}^A, \nohat{M}^{AB} \big\} \,,
\end{align}
with the 24 two-forms $\nohat{B}^A$ being
either self-dual or anti-self-dual.
There is no generalized SS reduction of IIB supergravity on K3 since
the IIB theory has only field strengths of odd rank while the manifold K3
has only harmonic forms of even rank.
Therefore the six-dimensional IIB pseudo--action \cite{Bergshoeff:1996sq}
has no mass parameters and reads
\begin{align}
  \mathcal{L}_{\text{IIB,6}} = & \tfrac{1}{2} \sqrt{-\nohat{g}}
e^{-2\nohat{\phi}}
    \Big[ -\nohat{R} + 4 |d \nohat{\phi}|^2 -\tfrac{3}{4}
|d \nohat{B}|^2
    +\tfrac{1}{8} \text{Tr}(d \nohat{M} d \nohat{M}^{-1})
    +\tfrac{1}{2} e^{2\nohat{\phi}} |d \nohat{\ell}^A|^{2}
    +\tfrac{3}{8} e^{2\nohat{\phi}} |\nohat{H}^A|^{2} \Big] \notag \\[2mm]
  & \qquad -\tfrac{1}{16} \nohat{\epsilon}
    \Big[ \nohat{\ell}_A d \nohat{B} d \nohat{B}^A
    \Big] \,,
\label{IIB6}
\end{align}
with the field strengths and the (anti-)self-duality relations
\begin{align}
  \nohat{H}^A \equiv d \nohat{B}^A + \nohat{\ell}^A
d \nohat{B} , \qquad
  \nohat{H}^A = L^{AB} \nohat{M}^{-1}_{BC}\, {}^\star \nohat{H}^C \,.
\end{align}
The action (\ref{IIB6}) has the following axionic shift symmetries
\begin{align}
  \nohat{\ell}^A \rightarrow \nohat{\ell}^A + m^A , \qquad
  \nohat{B}^A \rightarrow \nohat{B}^A - m^A \nohat{B} \,,
\label{shift}
\end{align}
with $m^A$ constant.
We will use this symmetry below to introduce masses when reducing to
five dimensions\footnote{
By using the full $O(5,21)$ U-duality symmetry one can
induce more mass
  parameters in five dimensions, as discussed in Appendix B.
In this Section we only use the axionic
  shift symmetries since this suffices for the derivation of the
massive T-duality rules.}.

To establish T-duality between the IIA and IIB theories an isometry is
 required.
As in the fluxless case, we will reduce both
IIA and IIB actions to five dimensions.
The IIA reduction formulae read (expressing 6D fields in terms of 5D fields)
\begin{align}
  \begin{array}{rcl}
  \underline{\text{iia}} & & \underline{\text{5D}}  \\
  {\nohat g}_{\underline{zz}} &=&  - e^{-2\phi -2\sigma/{\sqrt 3}}\, , \\
  {\nohat g}_{{\underline z}\mu} &=&  -e^{-2\phi -2\sigma/{\sqrt 3}}A_\mu\, , \\
  {\nohat g}_{\mu\nu} &=&  e^{-2\phi + 2\sigma/{\sqrt 3}} g_{\mu\nu}
    - e^{-2\phi -2\sigma/{\sqrt 3}}A_\mu A_\nu\, ,  \\
  {\nohat B}_{\mu\nu} &=&  (B+AC)_{\mu\nu}\, , \\
  {\nohat B}_{\underline z \mu} &=&  C_\mu\, , \\
  \nohat \phi &=&  -\phi + \tfrac{1}{{\sqrt 3}} \sigma\, , \\
  {\nohat V}_\mu^A &=&  V_\mu^A +\ell^A A_\mu\, , \\
  {\nohat V}_{\underline z}^A &=&  \ell^A\, , \\
  {\nohat M}^{AB} &=&  M^{AB} \, ,
  \end{array}
\label{IIAred}
\end{align}
where we use the notation $(AC)_{\mu\nu} = A_{[\mu}C_{\nu]}$.
These reduction rules are identical to the massless case, i.e.~they
correspond to ordinary Kaluza-Klein reductions.
There is no dependence on the internal coordinate $\underline{z}$.
Instead, the IIB reduction formulae read, again expressing 6D fields in
terms of 5D fields,
\begin{align}
  \begin{array}{rcl}
  \underline{\text{iib}} & & \underline{\text{5D}}   \\
  {\nohat g}_{\underline{zz}} &=& - e^{2\phi +2\sigma/{\sqrt 3}}\, , \\
  {\nohat g}_{{\underline z}\mu} &=& -e^{2\phi +2\sigma/{\sqrt 3}}C_\mu\, , \\
  {\nohat g}_{\mu\nu} &=& e^{-2\phi + 2\sigma/{\sqrt 3}} g_{\mu\nu}
    - e^{2\phi +2\sigma/{\sqrt 3}}C_\mu C_\nu\, , \\
  {\nohat B}_{\mu\nu} &=& (B+CA)_{\mu\nu} \,,  \\
  {\nohat B}_{\underline z \mu} &=& A_\mu\, , \\
  {\nohat B}_{\mu\nu}^A &=& (B^A + C V^A)_{\mu \nu} -2 m^A
(B+CA)_{\mu \nu} \underline{z}\, , \\
  {\nohat B}^A_{\underline z \mu} &=& V^A_\mu\, , \\
  {\nohat \phi} &=& \tfrac{2}{{\sqrt 3}} \sigma\, , \\
  {\nohat \ell}^A &=& \ell^A +2 m^A \underline{z}\, , \\
  {\nohat M}^{AB} &=& M^{AB} \, .
  \end{array}
\label{IIBred}
\end{align}
Note the linear dependence on the internal coordinate $\underline{z}$,
which takes the form of a  $\underline{z}$-dependent
axionic shift symmetry (\ref{shift}). This
particular $\underline{z}$-dependence will introduce
mass parameters in the five-dimensional theory. We are
dealing here with a Scherk-Schwarz reduction
similar to the one used in \cite{Bergshoeff:1996ui}.
For consistency in five dimensions, the shift symmetry \eqref{shift} is
crucial;
it implies that the reduced theory is independent of the
internal coordinate  $\underline{z}$.
After dimensional reduction, the (anti-)self-dual two-forms split
up into one- and two-forms.
The duality relations can be used to eliminate one of the two.
For obvious reasons we will keep the 24 vectors $V^A$ and eliminate the
two-forms $B^A$.
Reducing the above six-dimensional actions in this way, we find that both
the massive IIA {\it and} the IIB theory yield the following massive
five-dimensional $N=2$ theory:
\begin{align}
  \mathcal{L}_{\text{II,5}} = & \tfrac{1}{2} \sqrt{g} e^{-2\phi}
    \Big[ -R + 4 |d \phi|^2 -\tfrac{4}{3} |d \sigma|^2
    -\tfrac{3}{4} e^{4(\phi-\sigma/\sqrt{3})} |H|^2
    +\tfrac{1}{8} \text{Tr}(d M d M^{-1})
    - |F^A|^{2} + \notag \\[2mm]
  & \hspace{2.1cm} + \tfrac{1}{2} e^{4\sigma/\sqrt{3}} |G^A|^{2}
    - 2 e^{-4(\phi-\sigma/\sqrt{3})} |m^A|^{2}
    + e^{4\phi} |d C|^2
    + e^{-4\sigma/\sqrt{3}} |d A|^2 \Big] + \notag \\[2mm]
  & +\tfrac{1}{4} \epsilon \Big[ C F^A F_A + (B+CA) F^A G_A +
    2 m^A B C F_A - \tfrac{1}{2} m^A (B+CA)^2 d \ell_A \Big] \,,
\label{action5d}
\end{align}
with the field strengths
\begin{align}
  H = d B + C d A + A d C , \qquad
  F^A = d V^A + \ell^A d A + m^A (B+CA) , \qquad
  G^A = d \ell^A - 2 m^A C \,.
\label{5dfs}
\end{align}
Note that the mass parameters $m^A$ break the $O(5,21)$ symmetry to an
$O(5,21)$ pseudo-symmetry.  An $O(4,20)$ subgroup acts as
\begin{align}
  V^A \rightarrow \Omega^A_{\; B} V^B \,, \qquad
  M^{AB} \rightarrow \Omega^A_{\; C} \Omega^B_{\; D} M^{CD} \,, \qquad
  m^A \rightarrow \Omega^A_{\; B} m^B \,, \qquad
  \ell^A \rightarrow \Omega^A_{\; B} \ell^B \,,
\end{align}
with $\Omega^A_{\; B} \in O(4,20)$.
The fields $C$ and $B$ are massive, as could be expected from their
 six-dimensional IIA origin.

Having the relations \eqref{IIAred} and \eqref{IIBred} between six- and
five-dimensional fields at our disposal, we can use these to relate fields
in six dimensions with one isometry.
We thus derive the following
massive T-duality relations between Type IIA and Type IIB string
theory compactified on K3:
\begin{align}
  \begin{array}{rcl}
  \underline{\text{iib}} & & \underline{\text{iia}} \\
  {\nohat \phi} &=& \nohat \phi -\tfrac{1}{2}{\rm log}
(-{\nohat g}_{\underline {zz}})\, ,\\
  {\nohat g}_{\underline {zz}} &=& 1/{\nohat g}_{\underline {zz}}\, ,
\\
  {\nohat g}_{\underline z \mu} &=& {\nohat B}_{\underline z\mu}/
    {\nohat g}_{\underline {zz}}\, ,\\
  {\nohat g}_{\mu\nu} &=& {\nohat g}_{\mu\nu} -  (
    {\nohat g}_{\underline z\mu}{\nohat g}_{\underline z\nu}
    - {\nohat B}_{\underline z\mu}{\nohat B}_{\underline z\nu} )/
    {\nohat g}_{\underline {zz}}\, ,\\
  {\nohat B}_{\underline z\mu} &=& {\nohat g}_{\underline z\mu}/
    {\nohat g}_{\underline {zz}} \, ,\\
  {\nohat B}_{\mu\nu} &=& {\nohat B}_{\mu\nu}
    -  ({\nohat g}_{\underline z\mu} {\nohat B}_{\underline z\nu}
    - {\nohat g}_{\underline z\nu} {\nohat B}_{\underline z\mu} )
    / {\nohat g}_{\underline {zz}} \, ,\\
  {\nohat B}_{\underline z\mu}^A &=& {\nohat V}_\mu^A -
{\nohat V}_{\underline z}^A
    {\nohat g}_{\underline z\mu}/{\nohat g}_{\underline {zz}} \, ,\\
  {\nohat\ell}^A &=& {\nohat V}_{\underline z}^A + 2 m^A \underline{z}
\, ,\\
  {\nohat M}^{AB} &=& {\nohat M}^{AB}\, ,
  \end{array}
\label{mTrules}
\end{align}
where at the left-hand side are IIB fields
and at the right-hand side are IIA fields.
Note that the duality transformations of ${\nohat B}_{\mu\nu}^A$ are not
given.
Their duality rules are encoded in the reduced self-duality
conditions relating IIA and IIB field strengths:
\begin{align}
  \tfrac{3}{2} \nohat{H}^A = - e^{2\nohat{\phi}} \nohat{M}^{AB}\,
 {}^\star \nohat{F}_B \,.
\end{align}
In comparison with the massless case, only the T-duality rules for
$\nohat{\ell}^A$ and $\nohat{H}^A$ receive massive corrections.
The latter one receives corrections due to the fact that the curvature
$\nohat F^B$ in the above duality relation receives massive corrections.

This finishes our derivation of the massive T-duality rules.

\subsection{Brane Solutions}

Our starting point is the set of
basic dp-brane solutions\footnote{
To indicate the higher-dimensional origin of the
six-dimensional solutions we use lower-case
letters in the six-dimensional case and higher case letters in the
ten- or eleven-dimensional case. In Section 4
one can find the precise correspondence between the lower- and
higher-dimensional branes.} $(p=0,1,2,3,4)$
that preserves 1/2 supersymmetry.
The general dp-brane solution is given by \cite{Behrndt:1997pm}
\begin{align}
  ds^2_{dp} & = {\Omega}^{-1} \Big[ dt^2 - ds_p^2 \Big] -
{\Omega} dx_{5-p}^2\, , \notag \\
  e^{2 \phi} & = {\Omega}^{1-p}\, , \notag \\
  F^A_{01..pm} & = \partial_m H^A\, , \notag \\
  \mathcal{M}^{AB} & = \mathbb{I}^{AB} + 2 {\Omega}^{-1} H^A H^B\, ,
\label{dp-brane}
\end{align}
where $F^A$ is a (p+2)-form field strength and
${\Omega}^2 \equiv H_A H^A$. The functions $H^A$ are
harmonic in the (5--p)-dimensional transverse space.
The dp-branes can carry 24 different charges
$\forall p$:
1 x 2 = 2 come from Dp $\bot$ D(p+4) and 3 x 2 + 16 x 1 = 22
from D(p+2) $\bot$ D(p+2)
in terms of ten-dimensional intersections\footnote{
The total number of independent charges can also be deduced 
by counting the independent ways in which one can obtain a nonzero expression
for $m^A m^B L_{AB}$ thereby using the fact that
$L_{AB}$ has a $8 \times 8$ off-diagonal piece and a 16x16 diagonal piece.
Thus one obtains $4 \times 2 + 16 \times 1 = 24$ independent charges.}.
All these
intersections have an obvious 11D interpretation in terms of M-theory
compactified on $S^1\times K3$, i.e.~first compactifying on $S^1$
 and then
on $K3$. For $p=0, 2, 4$ we obtain $M5 // W1, M2 // KK6$ and
$KK6 // KK6$,
respectively. Here it is understood that the special
isometry of the KK6-branes is in the $S^1$-direction.

We note that the dp-branes with $p=1,3$ are solutions of the
iib theory which has no mass parameters. The other dp-branes,
with $p = 0,2,4$ are solutions of the iia theory which can have
massive deformations. Only the construction of the d4-brane
solution requires the presence of a mass parameter, as can be
seen from the expression for the harmonic functions $H^A$:
\begin{equation}
H^A = 1 + m^A x\, ,
\end{equation}
where $x$ is the single transverse direction. To stress this fact we
will call this solution the md4-brane (massive d4-brane). The
d0-brane and d2-brane solutions (\ref{dp-brane}) are only valid in the
massless case, i.e.~in the absence of any mass parameters.
To relate the d3-brane and md4-brane solutions
one must use the massive T-duality rules
\eqref{mTrules}. This is the natural replacement of the 10D massive
T-duality rules of \cite{Bergshoeff:1996ui}.
The reason that the massive T-duality works is that the
d3-brane solution, assuming an extra isometry direction in the
two-dimensional transverse space, has the proper (linear)
$\underline z$-dependence, i.e.~a linear axion.

The massive T-duality rules cannot be used to generate a massive md2-brane
solution out of the d3-brane. The reason is that in this case the
d3-brane must be reduced over a world-volume isometry direction
without any linear $\underline z$-dependence. Hence the reduction
does not introduces any mass parameters to relate to a massive
md2-brane solution. Instead, one can T-dualize the d3-brane to a
{\it massless} d2-brane solution.
Quite generally, one can only generate a massive
iia solution out of a massless iib solution if the iib solution
has the correct $\underline z$-dependence, i.e.~a linear axion.

Alternatively, we can apply the massive T-duality rules and construct 
a new iib solution, with a linear
axion, out of a given massive iia solution. We note that the iia theory
is a compactification of the massive
IIA supergravity theory for which massive BPS brane solutions
preserving 1/4 supersymmetry are known.
The simplest solution to consider is a Kaluza-Klein
compactication of the massive
(fundamental) string \cite{Janssen:1999sa, Massar:1999sb}.
This yields
the following D=6 massive fundamental string or
${\rm mf1}_{\rm iia}$-solution 
\begin{align}
& ds^2_{iia} = H^{-1} \Big[dt^2 - dz^2 \Big] - dx_4^2 \quad ,
\qquad e^{-2 \phi} = H\quad , \qquad V_t = m z
\notag \\
& B_{tz} = H^{-1}+1 \qquad , \qquad
H = m z - \sum_{r=1}^{r=4} \frac{m^2_r}{4} x_r^2 + H^\prime(x)\ .
\label{mstr}
\end{align}
Here $m$ and $V$ denote the $A=8$ component of $m^A$ and $V^A$, respectively.
The function $H^\prime(x)$ is a harmonic function in the 4 transverse
directions $x_r$ and $\sum_r m_r^2 = m^2$. The constants $m_r$
are integration constants. For $m=0$ the ${\rm mf1}_{\rm iia}$-solution
reduces to the  fundamental string or ${\rm f1}_{\rm iia}$-solution.
The ${\rm mf1}_{\rm iia}$-solution can be viewed as an intersection
of a f1-string ($B_{tz} \ne 0$), a d4-brane ($m \ne 0$) and a d0-brane
($V\ne 0$).

Note that $H$ depends on the
coordinate $z$ in the string direction.  Therefore, unlike in the
$m=0$ case, the string direction does not represent anymore an
isometry and we cannot T-dualize this direction to obtain a wave
solution of the iib theory. Interestingly enough,
there is a related iib wave solution \cite{Blau:2001ne}
which does not break any supersymmetry.  After T-dualizing the iib wave
solution of \cite{Blau:2001ne} back to the iia side,
the resulting metric and antisymmetric tensor are the same as for the
${\rm mf1}_{\rm iia}$-solution if one ignores the linear
$z$-dependence. Although this T-dual solution still solves the
equations of motion, supersymmetry is broken \cite{Blau:2001ne}.

Another possibility is to assume an isometry in one of the four transverse
directions, say x, and to T-dualize in this direction.
An application of the massive T-duality rules leads to a massive
${\rm mf1}_{\rm iib}$-solution with a linear axion:
\begin{align}
& ds^2_{iib} = H^{-1} \Big[dt^2 - dz^2 \Big] - dx_4^2 \quad ,
\qquad e^{-2 \phi} = H\quad , \qquad V_t = m z
\notag \\
& B_{tz} = H^{-1}+1 \qquad , \qquad
H = m z - \sum_{r=1}^{r=4} \frac{m^2_r}{4} x_r^2 + H^\prime(x)\ ,\notag\\
&\ell = 2mx\quad , \qquad \qquad C_{tx} = -mz\,.
\label{mstriib}
\end{align}
Here $\ell$ and $C_{tx}$ denote the $A=8$ component of $\ell^A$
and $B_{tx}^A$, respectively. The new ${\rm mf1}_{\rm iib}$-solution
can be viewed as an intersection of a f1-string ($B_{tz}\ne 0$),
a d3-brane ($\ell \ne 0$) and a d1-brane ($C_{tx}\ne 0$).

\section{Massive S-duality}

\subsection{Supergravity Relations}

We now discuss the massive S-duality.
In the massless case the S-duality we consider
relates the Heterotic string theory
compactified on $T^4$ with the Type IIA string theory
compactified on $K^3$. The explicit S-duality rules read
(with heterotic fields on the left-hand and iia fields on the right-hand side):
\begin{align}  
  \begin{array}{rcl}
  \underline{\text{h}} & & \underline{\text{iia}} \\
  g_{\mu \nu} & = & e^{-2 \phi} g_{\mu \nu} \,, \\
  \phi & = & -\phi \,, \\
  \tilde{H} & = & e^{-2 \phi}\, {}^\star H \,.
  \end{array}
\label{m0d}
\end{align}
Here we have used the definitions $\tilde H \equiv d \tilde B +
V_A dV^A$ and $H = dB$.

We have not been able to construct a massive S-duality between
the Heterotic string theory on $T^4$ and Type IIA string theory
compactified on K3. As has already been explained in
\cite{Haack:2001iz, Janssen:2001hy}, one of the obstructions to
construct a massive deformation of (\ref{m0d}) is that
the  NS 2-form at the IIA side, see eq.~(\ref{mfs}), becomes massive.
At first sight, one can therefore not apply a dualization procedure
for the 2-form in the Lagrangian. The situation
is similar to the massive 2-form of Romans supergravity.
{From} that case it is known that one can perform the
duality \eqref{m0d} also on massive 2-forms provided we dualize
the pair
\begin{equation}
\{B\ ({\rm massive}), m_A V^A\ ({\rm massless})\}\ \
 \longrightarrow\ \
\{\tilde B\ ({\rm massless}), m_A\tilde{V}^A\ ({\rm massive})\}\, .
\end{equation}
 Such
a massive S-duality can only be defined
at the level of the field equations \cite{Bergshoeff:1998ak}.
This leads to the following massive S-duality rules:
\begin{align}
  \begin{array}{rcl}
  \underline{\text{h}} & & \underline{\text{iia}} \\
  g_{\mu \nu} & = & e^{-2 \phi} g_{\mu \nu} \,, \\
  \phi & = & -\phi \,, \\
  \tilde H & = & e^{-2 \phi}\, {}^\star {H} \,, \\
  m_A \tilde{F}^A & = & m_A M^{AB}\, {}^\star F_B \,,
  \end{array}
\label{mAmassSdual}
\end{align}
with $m_A\tilde{F}^A$ a 4-form curvature.
The different (massive) curvatures are given by
\begin{align}
  \tilde{H} & = d \tilde{B} + V_A d V^A + 2 m_A V^A B - 2 m_A \tilde{V}^A
\,, \notag \\
  m_A \tilde{F}^A & = m_A d \tilde{V}^A - m_A V^A H
    + \tfrac{1}{2} m^2 B^2\,, \notag\\
H & = dB\,, \notag\\
F_A & = d V_A + m_A B \,,
\label{mAmasshet}
\end{align}
where $m_A\tilde{V}^A$ is a 3-form potential.
The fact that the S-dual version of the IIA theory (\ref{actionmiia})
does not allow an action shows that it has no obvious higher-dimensional
origin. Indeed, we have not been able to obtain the S-dual
theory via some generalised compactification of the 10D Heterotic
theory.

However, we are able to construct a massive S-duality between the
Heterotic string theory compactified on $T^4$ and M-theory
compactified on $K3 \times S^1$.  Our construction is based on the
observation that the Heterotic string theory compactified on $T^3$ is
S-dual to M-theory compactified on K3 \cite{Hull:1995ys,
Witten:1995ex}.  To construct the massive S-duality we apply a SS
reduction of both seven-dimensional theories on a circle $S^1$ to six
dimensions. This is in the spirit of the 4D massive dualities discussed
in \cite{Curio:2001ae}.

We first discuss the massless S-duality in seven dimensions
between the Heterotic string theory compactified on $T^3$
and M-theory on $K3$.
The field contents of the dual theories are
\begin{align}
  \text{Heterotic on } T^3: \qquad &
    \big\{ \noth{g}, \tilde{B}, \noth{\phi}, \noth{V}^a, \noth{M}^{-1}_{ab}
\big\} \,, \notag \\
  \text{M-theory on } K3: \qquad & \big\{ \noth{g}, \noth{C}^{(3)},
\noth{\phi},
    \noth{V}^a, \noth{M}^{-1}_{ab} \big\} \,,
\end{align}
where the scalar indices $a,b$ run from $1,\ldots,22$.
Note that the field content between the two theories only differs in the
sense that the Heterotic string on $T^3$ has a
two-form $\tilde{B}$ while M-theory on K3 contains a (dual)
three-form $\noth{C}^{(3)}$.
All other fields take the same form.
The scalar matrix $\noth{M}^{-1}_{ab}$ parametrises the coset
 $O(3,19)/(O(3)\times O(19)$.
It contains 57 scalars, of which 54 are axionic and 3 dilatonic.
The 22 vectors $\noth{V}^a$ fall into the fundamental representation of the
U-duality group.
The corresponding massless actions read
\begin{align}
  \mathcal{L}_{\text{het,7}} = & \tfrac{1}{2} \sqrt{g} e^{-2\phi}
    \Big[ -R + 4 |d \phi|^2
    -\tfrac{3}{4} |d \tilde{B} - V_a d V^a|^2
    +\tfrac{1}{8} \text{Tr} (d M d M^{-1})
    - |d V^a|^{2} \Big] \,, \notag \\[1mm]
  ~ \notag \\
  \mathcal{L}_{\text{m,7}} = & \tfrac{1}{2} \sqrt{g} e^{-2\phi}
    \Big[ -R + 4 |d \phi|^2 +
 \tfrac{1}{3} e^{4\phi/5} |d C^{(3)}|^2
    +\tfrac{1}{8} \text{Tr} (d M d M^{-1})
    - e^{8\phi/5} |d V^a|^{2} \Big] + \notag \\[2mm]
  & \qquad +\tfrac{1}{24} \epsilon C^{(3)} d V_a d V^a \,.
\end{align}
Both have a global $O(3,19)$ U-duality symmetry
\begin{align}
  V^a \rightarrow \Omega^a_{\; b} V^b \,, \qquad
  M^{ab} \rightarrow \Omega^a_{\; c} \Omega^b_{\; d} M^{cd} \,,
\label{HUduality}
\end{align}
with $\Omega^a_{\; b} \in O(3,19)$.
The two theories in 7D are related by the (massless) S-duality
\begin{align}
  \begin{array}{rcl}
  \underline{\text{7D h}} && \underline{\text{m}} \\
  g_{\mu \nu} &=& e^{-8 \phi/5} g_{\mu \nu} \,, \\
  \phi & = & -\phi \,, \\
  \tfrac{3}{4}\, {}^\star (d \tilde{B} - V_a d V^a)
  & = & e^{-2\phi} (d C^{(3)}) \,,
  \end{array}
\label{d7D}
\end{align}
with 7D heterotic fields on the left-hand side and M-theory fields on the
right-hand side.
Note that the third relation interchanges Bianchi identities and equations
of motion. Under this S-duality the Wess-Zumino term of the M-theory
is interchanged with
the Chern-Simons term in the 3-form field strength of the Heterotic theory.

We now perform a SS reduction on both seven-dimensional theories
using the results of Appendix B. The reduction rules for the
heterotic fields are given in Appendix B, see eq.~(\ref{red76}). The
reduction rules of the 7D M-theory fields are
\begin{equation}
  \begin{array}{rcl}
  \underline{\text{m}} && \underline{\text{iia}} \\
 {\nohat g}_{\underline{zz}} &=& - e^{8 \phi /5 -4 \sqrt 3 \sigma /5}\, ,\\
 {\nohat g}_{{\underline z}\mu}
&=& -e^{8 \phi /5 -4 \sqrt 3 \sigma /5}V^\bullet_\mu\, ,\\
  {\nohat g}_{\mu\nu} &=& e^{-2 \phi /5 + 8 \sqrt{3} \sigma / 15} g_{\mu\nu}
    - e^{8 \phi /5 -4 \sqrt 3 \sigma /5} V^\bullet_\mu V^\bullet_\nu\, ,\\
  C^{(3)}_{\mu\nu\rho} &=& C^{(3)}_{\mu\nu\rho} \, ,  \\
  C^{(3)}_{\mu\nu \underline z} &=& B_{\mu \nu}\, ,  \\
  \nohat \phi &=& \phi + \tfrac{1}{{\sqrt 3}} \sigma\, ,\\
  {\nohat V}_\mu^a &=& \Omega^{a}_{\; b} (z) (V_\mu^b +
    \ell^b V^\bullet_\mu) \, ,\\
  {\nohat V}_{\underline z}^a &=& \Omega^{a}_{\; b} (z) \ell^b \, ,\\
  {\nohat M}^{ab} &=& \Omega^{a}_{\; c} (z) \Omega^{b}_{\; d} (z) M^{cd}\, ,
  \end{array}
\label{Mred76}
\end{equation}
with $\Omega^{a}_{\; b} (z) \in O(3,19)$.
After reduction, the two six-dimensional theories have the
following field content:
\begin{align}
  \big\{ g, B, \phi, V^A, M^{AB} \big\} \,,
\end{align}
where the $A,B$ indices refer to the $O(4,20)$ U-duality group.
Here the three-form $C^{(3)}$ is dualised into a vector $V_\bullet$ via
$d C^{(3)} = \tfrac{1}{4} {}^* F_\bullet$. The O$(4,20)$ index A
and the $O(3,19)$ index $a$ are related via $A = \{{}^\bullet, {}_\bullet, a\}$,
see also Appendix B.
The 6D Lagrangians for the massive heterotic and iia actions read
\begin{align}
  \mathcal{L}_{\text{het},6} = \tfrac{1}{2} \sqrt{-g} e^{-2\phi}
    \Big[ & -R + 4 |d \phi|^2 -\tfrac{3}{4} |\tilde{H}|^2
    +\tfrac{1}{8} \text{Tr} (\mathcal{D} M \mathcal{D} M^{-1})
    - |F^A|^{2} + \notag \\[1mm]
  & - \tfrac{1}{12} f_{ABC} f_{DEF} M^{AD} (M^{BE} M^{CF} - 3 L^{BE} L^{CF})
 \Big] \,, \notag \\[1mm]
  ~ \notag \\
  \mathcal{L}_{\text{iia,6}} = \tfrac{1}{2} \sqrt{-g} e^{-2\phi}
    \Big[ & -R + 4 |d \phi|^2 -\tfrac{3}{4} |H|^2
    +\tfrac{1}{8} \text{Tr} (\mathcal{D} M \mathcal{D} M^{-1})
    - e^{2 \phi} |F^A|^{2} + \notag \\[1mm]
  & - \tfrac{1}{12} f_{ABC} f_{DEF} M^{AD} (M^{BE} M^{CF} - 3 L^{BE} L^{CF})
\Big]
    - \tfrac{1}{8} \epsilon B F_A F^A \,,
\end{align}
with the field strengths
\begin{align}
  H & = d B \,, \notag \\
  \tilde{H} & = d B - V_A d V^A - \tfrac{1}{6} f_{ABC} V^A V^B
V^C \,, \notag \\
  F^A & = d V^A - \tfrac{1}{2} f_{BC}{}^A V^B V^C \,, \notag \\
  \mathcal{D} M^{AB} & = d M^{AB} - f_{CD}{}^A V^C M^{BD} - f_{CD}{}^B
V^C M^{AD} \,.
\end{align}
The only non-zero structure coefficients are $f_{\bullet\,ab}=m_{ab}$
with $m_{ab}$ an anti-symmetric 22x22 matrix and $a,b$ indices of $O(3,19)$.
The Jacobi identity of $f_{ABC}$ is identically satisfied for
arbitrary $m_{ab}$.
Both theories have an $O(4,20)$ U-duality pseudo-symmetry
\begin{align}
  V^A \rightarrow \Omega^A_{\; B} V^B \,, \qquad
  M^{AB} \rightarrow \Omega^A_{\; C} \Omega^B_{\; D} M^{CD} \,, \qquad
  f^{ABC} \rightarrow \Omega^A_{\; D} \Omega^B_{\; E} \Omega^C_{\; F}
    f^{DEF} \,,
\end{align}
with $\Omega^A_{\; B} \in O(4,20)$.

We are now in a position to establish an S-duality between the two reduced
theories.
The massless S-duality \cite{Behrndt:1996si} maps the massless field
strengths $H$ and the Hodge dual $\tilde{H}$ into each other.
There is no obstruction to extend this to the massive case since
both $B$ and $\tilde B$ are massless. The massive
S-duality simply relates the massive field strengths to each other:
\begin{align}  
  \begin{array}{rcl}
  \underline{\text{h}} & & \underline{\text{iia}} \\
  g_{\mu \nu} & = & e^{-2 \phi} g_{\mu \nu} \,, \\
  \phi & = & -\phi \,, \\
  \tilde{H} & = & e^{-2 \phi}\, {}^\star H \,.
  \end{array}
\end{align}
with heterotic fields on the left-hand and iia fields on the right-hand side.
These relations are obtained from the 7D duality relations \eqref{d7D}
by the generalised
reduction relations \eqref{red76} of Appendix B
and \eqref{Mred76}. Thus they relate the two dual
massive theories in six dimensions.

This finishes our discussion of the massive S-duality rules.

\subsection{Brane Solutions}

The basic brane solutions of the Heterotic theory are the
h$p$-brane solutions with $p=0,2,4$. Only the h4-brane
solution requires a mass parameter. The h0-brane
and h2-brane solutions are given by (p=0,2):
\begin{align}
  ds^2_{hp} & = {H}^{p-2} \Big[ dt^2 - ds_p^2 \Big] -
    {H}^{p} dx_{5-p}^2\, , \notag \\
  e^{2 \phi} & = {H}^{p-1}\, , \notag \\
  F^A_{01..pm} & = \partial_m H^A\, , \notag \\
  \mathcal{M}^{AB} & = \mathbb{I}^{AB} + 2 {H}^{-1} H^A H^B\, .
\end{align}
These are S-dual to d$p$-brane solutions on the iia side and can be
obtained by
applying the massless S-duality rules \eqref{m0d} to the iia d0- and
 d2-brane solutions
\eqref{dp-brane}. The h0-branes are reductions of the electric chiral
null model $F1 // W1$,
which give (4,4) charges, and H0-branes carrying (0,16) electric charges
with respect to the
YM sector \cite{Behrndt:1997pm, Behrndt:1999mk}. The h2-branes are
reductions of the
magnetic chiral null model $S5 // KK5$, giving (4,4) charges, and the
magnetic H6 branes,
giving the (0,16) charges.

Upon applying the massive S-duality \eqref{mAmassSdual} to the d4-branes
of the iia theory, we obtain
${\tilde {h4}}$
domain wall solutions of the heterotic theory with $m^A$ deformations
\eqref{mAmasshet}.
However, as explained in the previous subsection, we have not been able
to find a higher-dimensional
origin of these massive deformations in the heterotic theory. Therefore
 we do not give the
explicit form of these solutions. By convention we will indicate
solutions for which there is no viable higher-dimensional
origin with a tilde.

On the other hand the mass deformations $f_{ABC}$ of the Heterotic theory,
which do have a clear higher-dimensional interpretation, 
give rise to the following h4-brane
solutions:
\begin{align}
  ds^2_{h4} & = \Big[ dt^2 - ds_4^2 \Big] - {H}^2 dx^2\, , \notag \\
  e^{2 \phi} & = {H}^{-1}\, , \notag \\
  \mathcal{M}_{ab} & \not= 0\, .
\label{h4-brane}
\end{align}
We have not worked out which components of $f_{ABC}$ support the h4
domain walls.
These correspond to reductions of the magnetic chiral null model and the
magnetic H6 branes. Since S-duality
can be extended to mass deformations $f_{ABC}$, the h4-branes have S-dual
partners in M-theory on K3 $\times$ $S^1$: these are m4-branes with
field configuration given by
\begin{align}
  ds^2_{m4} & = {H} \Big[ dt^2 - ds_4^2 \Big] - {H}^3 dx^2\, ,
\notag \\
  e^{2 \phi} & = {H}\, , \notag \\
  \mathcal{M}_{ab} & \not= 0\, .
\end{align}
Again, we have not worked out which components of $f_{ABC}$ support
the m4-brane solutions.
In the next Section we will discuss how the m4-branes can be uplifted to
the 11D $KK6 \bot KK6$ intersections.

\section{Higher Dimensional Origin}

We wish to consider the effect of S-duality on 1/2 BPS branes in 6D by
examining their
higher-dimensional origin. For this purpose
it is useful to express the 6D S-duality in terms of 10D S- and
T-dualities\footnote{
  Here we use that the different $\mathbb{Z}_2$-symmetries of string theory transform into
  each other under 10D S- and T-dualities \cite{Dabholkar:1997zd}.
For example:
  $(-)^{F_L} = S \, \Omega \, S^{-1}$.
We consider the modding out with the $\mathbb{Z}_2$-symmetries
only at the supergravity level. For a discussion of the twisted
sector, see \cite{Bergshoeff:1998re}.}:
\begin{alignat}{2}
  & \text{IIA on } T^4/ I_{6789} \equiv && \hspace{1cm}
\bf{\text{\bf IIA on } T^4/ I_{6789}
    \simeq K3} \notag \\
  & \updownarrow T_6 \notag \\
  & \text{IIB on } T^4/(-)^{F_L} I_{6789} \notag \\
  & \updownarrow S && \hspace{1.2cm} \updownarrow T_6~S~T_6 \notag \\
  & \text{IIB on } T^4/ \Omega I_{6789} \notag \\
  & \updownarrow T_6 \notag \\
  & \text{IIA on } T^4 / (-)^{F_L} I_{789} \Omega \equiv
    && \hspace{1cm} \bf{\text{\bf M-theory on } T^5 / I_{78911}
\simeq K3 \times S^1} \notag \\
  & \updownarrow T_{789} \notag \\
  & \text{IIB on } T^4/ \Omega && \hspace{1.2cm} \updownarrow S ~ T_{789}
\notag \\
  & \updownarrow S \notag \\
  & \text{IIB on } T^4 / (-)^{F_L} \equiv && \hspace{1cm} {\bf
    \text{\bf Heterotic on } T^4}
\label{10d6ddual}
\end{alignat}
\noindent
Here $T_x$ ($I_x$) denotes a T-duality (inversion) in the $x$-direction.
Note that in the case of the IIA-theory (M-theory) the K3 manifold lies
in the 6789 (78911)-direction. In the latter case we
have used for this interpretation that $(-)^{F_L} = I_{11} \Omega$.
{From} the Figure above we see how
the Heterotic model compactified on $T^4$ with enhanced gauge symmetry
is related, via duality, to M-theory compactified on $K3 \times S^1$ and
Type IIA string theory on
K3 in the singular orbifold limit.
Thus the 6D string theories are related by an $S~T_{789}$ duality
between Heterotic and M-theory and a $T_6~S~T_6$ duality between M-theory
and the Type IIA superstring.
These relations are conjectured for the full string theories, including
the massive cases
with background fluxes in the internal manifold.

At the supergravity level the relations correspond to the ordinary
6D duality relations in the massless case.
However, we have seen that the supergravities can have different massive
deformations,
depending on the order of the SS reductions. Sofar, we found two classes
of massive deformations: one with $m^A$ and one with $f_{ABC}$.
Therefore, one does not expect
that the massive deformations can be mapped onto each other by
$S~T_{789}$ and $T_6~S~T_6$ at the level of the supergravity theory.
We already concluded in Section 3 that the
mass deformations $m^A$ and $f_{ABC}$ do not have a higher-dimensional
origin in {\it all} supergravities.
In this section we will analyse this higher-dimensional
origin further by applying string dualities to brane intersections.

For our present purpose it is sufficient to consider a truncation of the 6D
theories. The full theories have a supergravity multiplet (with 4 
gravi-photons) coupled to 20 vector multiplets. We consider the truncation
to 4 vector multiplets. This corresponds to the reduction of D=10, N=1
supergravity on $T^4$ i.e.~Heterotic theory without Yang-Mills sector,
IIA on $T^4 / I_{6789}$ i.e.~the untwisted sector of
IIA string theory on K3 or M-theory on $T^5 / I_{78911}$
i.e.~the untwisted
sector of M-theory on K3 $\times S^1$. {From} now on we will always
refer to these
truncated theories which have an $O(4,4)$ duality symmetry.

We now apply the translation \eqref{10d6ddual} of 6D duality to
10D duality to
the brane solutions.
We will use the lower-case letters h, m and d to denote solutions of
Heterotic on $T^4$,
M-theory on K3 $\times S^1$ and IIA on K3 respectively.
We have summarized some of the nomenclature in the Table below. 
From a 6D point of view, the iia and h theories
with $f_{ABC}$-deformations are related via a S-duality. The two
iia theories with $m^A$- and $f_{ABC}$-deformations cannot be related 
via a 6D supergravity relation (in the massless case
they are related via an $O(4,20)$ rotation). The higher-dimensional origin
suggests a relation via a $T_6 S T_6$ string duality but sofar we
are unable to realize this 10D string duality,
after compactification, by a 6D supergravity duality.
Note that
not all components of $f_{ABC}$ contain mass parameters. To be
specific, we only have $f_{\bullet\,ab}=m_{ab}$.

\begin{center}
\begin{tabular}{|c|c|c|c|}
\hline
Theory&Deformation&Origin&Branes\\
\hline
iia&$m^A$&IIA on K3&dp\\
iia&$f_{ABC}$&M on K3$\times S^1$&mp\\
h&$f_{ABC}$&H on $T^4$&hp\\
\hline
\end{tabular}
\end{center}
\noindent{\it Table 1:\ the Table indicates the massive deformations
  of the different 6D supergravity theories, their higher-dimensional origin
  and the nomenclature for their brane solutions.}
\bigskip

In the remaining part of this Section we will only consider solutions 
that preserve 1/2 supersymmetry.
We therefore have to resort to intersections in the
higher-dimensional theory. We consider 5 different cases.

\begin{description}
\item{(1)}\ 
We first consider the 0-branes. 
We find that the S-duality in 6 dimensions relates, when uplifted to 
ten dimensions,  the following 
10D intersections\footnote{
  Here the notation is as follows: $\x$ corresponds to a worldvolume 
direction, $-$ to a
  transverse direction and $z$ to a Killing direction. The latter 
possibility corresponds
  to a $R^p$ radius dependence of the effective tension with 
$p>1$ \cite{Lozano-Tellechea:2000mc}.}:
\begin{eqnarray}
  d0 ~ \leftrightarrow &  ~ (0 | D2,D2): & ~
  \mbox{ {\scriptsize
        $\left\{ \begin{array}{c|ccccccccc}
         \x & - & - & - & - & - &  - & \x & \x &  - \\
         \x & - & - & - & - & - & \x &  - &  - & \x \\
                         \end{array} \right.$ } }\notag  \\
  && \hspace{-2cm} \updownarrow \notag \\
  m0 ~ \leftrightarrow &  ~ (0 | M2,M2): & ~
  \mbox{ {\scriptsize
        $\left\{ \begin{array}{c|cccccccccc}
         \x & - & - & - & - & - & - & \x & \x &  - & - \\
         \x & - & - & - & - & - & - &  - &  - & \x & \x  \\
                         \end{array} \right.$ } } \notag \\
  && \hspace{-2cm} \updownarrow \notag \\
  h0 ~ \leftrightarrow &  ~ (1 | F1,W1): & ~
  \mbox{ {\scriptsize
        $\left\{ \begin{array}{c|ccccccccc}
         \x &  - & - & - & - & - & - & - & -  & \x \\
         \x &  - & - & - & - & - & - & - & -  &  z \\
                         \end{array} \right.$ } }
\end{eqnarray}
We find the electric chiral null model at the heterotic side.
The duality between the electric chiral null model and intersections of
D-branes was already
noted in \cite{Behrndt:1997pm}. The D2 $\bot$ D2 intersections
given above yield six charges
while the other
two charges come from D0 // D4, which we do not give here.
By considering the orbifold limit we can easily deduce
the M-theory origin of the m0-branes. Six come from M2 $\bot$ M2 
while the other two come
from $W1 // M5$.
Note that since $T_6~S~T_6$ is a symmetry of massless IIA/M-theory
compactified to 6 dimensions,
the d0- and m0-branes constitute the same set of eight
0-branes and have the same
eleven-dimensional origin.

\item{(2)}\ 
We next consider the 2-branes. Six of them arise as
\begin{eqnarray}
  d2 ~ \leftrightarrow &  ~ (2 | D4,D4): & ~
  \mbox{ {\scriptsize
        $\left\{ \begin{array}{c|cccccccccc}
         \x & \x & \x & - & - & - & \x &  - &  - & \x \\
         \x & \x & \x & - & - & - &  - & \x & \x &  -
                         \end{array} \right.$ } } \notag \\
  && \hspace{-2cm} \updownarrow \notag \\
  m2 ~ \leftrightarrow &  ~ (3 | M5,M5): & ~
  \mbox{ {\scriptsize
        $\left\{ \begin{array}{c|cccccccccc}
         \x & \x & \x & - & - & - & \x &  - &  - & \x & \x \\
         \x & \x & \x & - & - & - & \x & \x & \x &  - &  -
                         \end{array} \right.$ } } \notag \\
  && \hspace{-2cm} \updownarrow \notag \\
  h2 ~ \leftrightarrow &  ~ (5 | S5,KK5): & ~
  \mbox{ {\scriptsize
        $\left\{ \begin{array}{c|cccccccccc}
         \x & \x & \x & - & - & - & \x & \x & \x &  - \\
         \x & \x & \x & - & - & - & \x & \x & \x &  z
                         \end{array} \right.$ } }
\end{eqnarray}
We find the magnetic chiral null model at the Heterotic side
\cite{Behrndt:1997pm, Behrndt:1996kr}.  The other two 2-branes
come from D2 // D6 and  M2 // KK6, respectively.  The d2- and
m2-branes are related by an element of the $O(4,4)$ duality symmetry
group.

\item{(3)}\ 
We now consider (some of) the 4-branes. They are of course the
most interesting from the point of view
of this paper. As we will demonstrate, not all domain
walls  have dual partners with a
viable higher-dimensional interpretation.
In Section 3 we derived
a supergravity correspondence between the domain walls of
Heterotic on $T^4$ and M-theory on
K3 $\times S^1$ with mass parameters contained in a
3-rank tensor $f_{ABC}$. This is the massive S-duality between
the h4-brane and m4-brane which is diagrammatically described
as follows:

\begin{eqnarray}
  \tilde{d4} ~ \leftrightarrow &  ~ (4 | KK6,KK6): & ~
  \mbox{ {\scriptsize
        $\left\{ \begin{array}{c|cccccccccc}
         \x & \x & \x & \x & \x & - &  z & \x &  - & \x \\
         \x & \x & \x & \x & \x & - & \x &  z & \x &  -
                         \end{array} \right.$ } } \notag  \\
  && \hspace{-2cm} \updownarrow \notag \\
  m4 ~ \leftrightarrow &  ~ (4 | KK6,KK6): & ~
  \mbox{ {\scriptsize
        $\left\{ \begin{array}{c|cccccccccc}
         \x & \x & \x & \x & \x & - & - & \x &  - & \x &  z \\
         \x & \x & \x & \x & \x & - & - &  z & \x &  - & \x
                         \end{array} \right.$ } } \notag  \\
  && \hspace{-2cm} \updownarrow \notag \\
  h4 ~ \leftrightarrow &  ~ (5 | S5,KK5): & ~
  \mbox{ {\scriptsize
        $\left\{ \begin{array}{c|cccccccccc}
         \x & \x & \x & \x & \x & - & - & - & \x & - \\
         \x & \x & \x & \x & \x & - & - & - & \x & z
                         \end{array} \right.$ } }
\end{eqnarray}
We have also indicated a $\tilde {d4}$-brane at the iia side. We have put
a tilde to indicate that this brane has no viable higher-dimensional
interpretation.
In the present case $\tilde {d4}$ is a solution of the IIA-theory
compactified
on K3 with $f_{ABC}$-deformations whose D=6 supergravity formulation
is unknown, in the sense that we do not know how to obtain such a theory
from SS-reduction.

With the above 4-branes we can connect to all the mass parameters 
coming from axionic shift symmetries.
In particular, the theories considered in Section 3 obtained mass 
parameters using the U-duality
in 7D. In the truncated case this is $O(3,3)$, having 6 axions. At the
heterotic side this
 corresponds to choosing the
worldvolume direction of the Heterotic magnetic chiral null model 
(with two charges) in one of the three
directions of $T^3$. The 4-brane in 7D becomes a domain wall upon 
reduction on a circle and is
therefore exactly the solution carried by $f_{ABC}$.
On the M-theory side, the m$4$-branes correspond
to two Kaluza-Klein monopoles wrapping different 2-cycles of the K3. 
Their isometries
always lie in the K3 and not in the $S^1$. Finally, in terms of IIA on K3 the
higher-dimensional origin of the $\tilde{{\rm d}4}$-branes lies in 10D 
KK6-branes. 
These 10D KK6-branes are reductions of 11D KK6-branes 
in a transverse coordinate. 
We do not know
how to use these 10D KK6-branes
for SS reduction of supergravity, as was found in 
Section 3.

\item{(4)}\ 
We next 
consider the 1/2 BPS 4-brane corresponding to the mass
deformations $m^A$ of IIA on K3. Its higher-dimensional origin is, as
for the other dp-branes,  an intersection of two D-branes. The $D6 \bot D6$
gives rise to six masses (in the truncated theory) while the other two
come from
$D4 // D8$. One should not expect to find an obvious interpretation of
these
deformations on the M-theory and Heterotic sides, as we have learnt in
Section 3.
Applying the 10D duality scheme to the first intersection we find
\begin{eqnarray}
  d4 ~ \leftrightarrow & ~ (4 | D6,D6): & ~
  \mbox{ {\scriptsize
        $\left\{ \begin{array}{c|cccccccccc}
         \x & \x & \x & \x & \x & - & \x & \x &  - &  - \\
         \x & \x & \x & \x & \x & - &  - &  - & \x & \x
                         \end{array} \right.$ } } \notag  \\
  && \hspace{-2cm} \updownarrow \notag \\
  \tilde{m4} ~ \leftrightarrow &  ~ (4 | KK6,KK6): & ~
  \mbox{ {\scriptsize
        $\left\{ \begin{array}{c|cccccccccc}
         \x & \x & \x & \x & \x & - & z & \x &  - &  - & \x \\
         \x & \x & \x & \x & \x & - & z &  - & \x & \x &  -
                         \end{array} \right.$ } } \notag  \\
  && \hspace{-2cm} \updownarrow \notag \\
  \tilde{h4} ~ \leftrightarrow &  ~ (4 | ??,??): & ~
  \mbox{ {\scriptsize
        $\left\{ \begin{array}{c|cccccccccc}
         \x & \x & \x & \x & \x & - & z & z & z & z \\
         \x & \x & \x & \x & \x & - & z & z & z & z
                         \end{array} \right.$ } }
\end{eqnarray}
In terms of M-theory
 on  K3 $\times S^1$, the $\tilde{{\rm m}4}$-brane is an intersection
of KK6-branes
with their isometry in the $S^1$. Note that this would be an
exotic brane in 7D, where its transverse space would be two-dimensional
 including an
isometry. We do not know how to generate the masses
corresponding to the $\tilde{m4}$-branes in the SS reduction of
M-theory on K3 $\times S^1$.
Similarly, in 
terms of the Heterotic theory compactified on $T^4$ the higher-dimensional
interpretation is absent, in agreement with the supergravity findings
of Section 3.

\item{(5)}\ 
The other 10D intersection reducing to a d4-brane is $D4 // D8$, which gives
\begin{eqnarray}
  d4 ~ \leftrightarrow & ~ (4 | D4,D8): & ~
  \mbox{ {\scriptsize
        $\left\{ \begin{array}{c|cccccccccc}
         \x & \x & \x & \x & \x & - &  - &  - &  - &  - \\
         \x & \x & \x & \x & \x & - & \x & \x & \x & \x
                         \end{array} \right.$ } } \notag \\
  && \hspace{-2cm} \updownarrow \notag \\
  \tilde{m4} ~ \leftrightarrow &  ~ (5 | M5,M9): & ~
  \mbox{ {\scriptsize
        $\left\{ \begin{array}{c|cccccccccc}
         \x & \x & \x & \x & \x & - & \x &  - &  - &  - &  - \\
         \x & \x & \x & \x & \x & - &  z & \x & \x & \x & \x
                         \end{array} \right.$ } } \notag  \\
  && \hspace{-2cm} \updownarrow \notag \\
 \tilde{h4} ~ \leftrightarrow &  ~ (4 | ??,??): & ~
  \mbox{ {\scriptsize
        $\left\{ \begin{array}{c|cccccccccc}
         \x & \x & \x & \x & \x & - & \x & z & z & z \\
         \x & \x & \x & \x & \x & - &  z & z & z & z
                         \end{array} \right.$ } }
\label{M5M9}
\end{eqnarray}
In terms of M-theory
 on K3 $\times S^1$ the $\tilde{{\rm m}4}$-brane is
an M5//M9 intersection. The M9 brane is not a solution of the
known 11D supergravity theory.
At the Heterotic side we cannot find a viable higher dimensional
interpretation in terms of a known intersection, as is indicated
by the question marks.
\end{description}
\bigskip

One might expect that there are other heterotic theories in 6D by
performing the Horava-Witten
scenario \cite{Horava:1995qa} in a direction different than the 11th.
 One could for example consider M-theory on
$T^4 \times (S^1 / \mathbb{Z}_2)$. This is related to
the Heterotic theory on $T^4$ by
\begin{alignat}{2}
  & \text{IIB on } T^4/ (-)^{F_L} \equiv && \hspace{1cm}
\bf{\text{\bf H on } T^4 }\notag \\
  & \updownarrow S \notag \\
  & \text{IIB on } T^4/ \Omega \notag && \hspace{1.2cm}
\updownarrow S~T_6 \\
  & \updownarrow T_6 \notag \\
  & \text{IIA on } T^4/ I_{6} \Omega \equiv && \hspace{1cm}
\bf{\text{\bf M-theory on }
    T^4 \times (S^1 / \mathbb{Z}_2)}
\end{alignat}
By applying a $S~T_6$ transformation
one can see that the massive deformations of
the above two theories
can be mapped onto each other\footnote{Up to a $T_6$
  transformation which acts as an element
  of $O(4,4)$ and therefore only rotates the $f_{ABC}$'s.}.
It is impossible to relate these masses to
the $m_A$-deformations of IIA on K3.

In summary, the analysis of the higher-dimensional origin partially
resolves the puzzle of massive
S-duality. The h4- and m4-branes map onto each other under
the massive S-duality map, as we also found in the supergravity approach.
The S-duals of the d4-branes, on
the other hand, do not have a viable 10D brane interpretion at the
Heterotic side.
This analysis also tells us about the missing mass parameters
in the theories. It is intriguing that both the d4- and the m4-branes
and their $T_6~S~T_6$ duals correspond to $KK6 \bot KK6$ intersections
in 11D (apart from the
$M5 // M9$ intersection). Thus one has a brane interpretation for both
 the $m_A$ and $f_{ABC}$
deformations in both theories. However, the corresponding mass parameter
only
appears in the 6D supergravity when the isometry of the KK monopole lies
 in the first of the
two compactification manifolds. Explicitly, the isometry in the $S^1$
corresponds to the $m_A$
deformations and these are only found in IIA on K3 or M on $S^1 \times$
K3. When the isometry
lies in the K3 this corresponds to the $f_{ABC}$'s and these only occur
in M theory on K3 $\times S^1$.

\section{Discussion and Summary}

Using a generalized SS reduction scheme
we have found two families of massive
deformations in six dimensional supergravities as illustrated in Figure 2.
In one case the mass parameters are in the 
fundamental representation of the U-duality
group O(4,20), in the other case they are in the 
3-index antisymmetric representation of $O(4,20)$. 
The T-duality between IIA and IIB on K3
can be extended with 24 $m_A$-deformations while the S-duality between
11D supergravity on K3 $\times S^1$ and the Heterotic theory
on $T^4$ can be extended
with 231
$f_{ABC}$-deformations. Thus, in this work we have found that there are
at least two cases where dualities between six-dimensional supergravities can
include mass parameters.

\begin{figure}[h]

\begin{picture}(400,220)(-100,-10)

\put(100,200){\makebox(0,0){\large \bf M}}
\put(-50,150){\makebox(0,0){\large \bf H}}
\put(200,150){\makebox(0,0){\large \bf IIA$_1$}}
\put(300,150){\makebox(0,0){\large \bf IIB}}
\put(150,50){\makebox(0,0){\large \bf m$_1$}}
\put(50,50){\makebox(0,0){\large \bf m}}
\put(-50,0){\makebox(0,0){\large \bf h$_{231}$}}
\put(200,0){\makebox(0,0){\large \bf iia$_{24}$}}
\put(50,0){\makebox(0,0){\large \bf iia$_{231}$}}
\put(155,0){\makebox(0,0){\large \bf iia$_{1} \, \in$}}
\put(300,0){\makebox(0,0){\large \bf iib}}
\put(90,190){\line(-4,-1){130}}\put(25,190){\makebox(0,0){\large
$S^1/\mathbb{Z}_2$}}
\put(150,40){\line(0,-1){30}}\put(165,25){\makebox(0,0){\large $S^1$}}
\put(50,40){\line(0,-1){30}}\put(65,25){\makebox(0,0){\large $S^1$}}
\put(-50,140){\line(0,-1){130}}\put(-65,75){\makebox(0,0){\large $T^4$}}
\put(200,140){\line(0,-1){130}}\put(185,75){\makebox(0,0){\large $K3$}}
\put(300,140){\line(0,-1){130}}\put(285,75){\makebox(0,0){\large $K3$}}
\put(105,190){\line(1,-3){42}}\put(140,140){\makebox(0,0){\large $K3$}}
\put(95,190){\line(-1,-3){42}}\put(60,140){\makebox(0,0){\large $K3$}}

\put(0,0){\vector(-1,0){30}\vector(1,0){60}\makebox(-60,-17){\large
${\cal S}_{231}$}}
\put(0,50){\vector(-1,0){30}\vector(1,0){60}\makebox(-60,-17){\large
${\cal S}$}}
\put(250,0){\vector(-1,0){30}\vector(1,0){60}\makebox(-60,-17){\large
${\cal T}_{24}$}}
\put(250,150){\vector(-1,0){30}\vector(1,0){60}\makebox(-60,17){\large
${\cal T}_1$}}

\end{picture}

\it \caption{Web of massive dualities.
  The vertical lines are SS reductions and the
  subscripts are the number of massive deformations.
  Note that there are two families
  of massive deformations in six dimensions (24 $m^A$-deformations
  and 231 $f_{ABC}$-deformations, see the text).}

\end{figure}
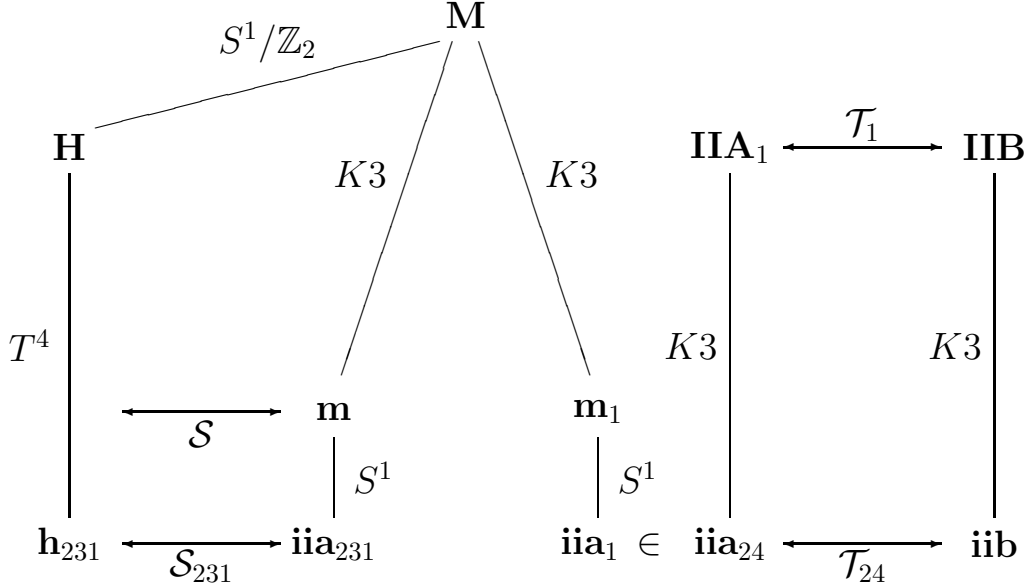

In Section 2 we have demonstrated the massive T-duality in
six dimensions between Type IIA string theory compactified
on K3 with background fluxes and Type IIB string theory
compactified on K3\footnote{
  One can reproduce one of the 24 $m^A$'s from M-theory on 
  K3$\times S^1$ as shown in Figure 2 \cite{Lu:1997rh, Haack:2001iz}.
  This corresponds to the $T_6~S~T_6$-relation of the D4- and M5-brane
  \eqref{M5M9}.}. 
Thus these theories
are equivalent in five dimensions.
In this paper we used the simplest scenario where
one includes background fluxes for
24 of the 100 axions in the 6D iib theory,
and no background flux at the iia side. 
Instead, one can use the generalised reduction
scheme of Appendix B
for the SS compactification of the iib theory from 6 to 5 dimensions
and this induces more mass parameters
\cite{Meessen:1999qm}. A subset of these parameters is
contained in the massive T-duality rules given in this paper.

As a by-product, see Appendix B,
we worked out the generalised toroidal SS reduction of
the Heterotic theory. At every step in this reduction 
we employ the full U-duality group. In this way one
obtains more mass parameters than if one only makes use
of the axionic shift symmetries.
In the case of a $T^4$-reduction the total
number of mass parameters is $120 + 153 + 190
+231 = 694$, which fit in a $f_{ABC}$ multiplet and are constrained by
the Jacobi identity.
A subset of these, $0+16+34+54=104$, are generated by axionic shift
symmetries and were
derived in \cite{Kaloper:1999yr}. It remains to
be checked whether our generalised SS reduction scheme gives rise to more
independent mass parameters than those which follow from the
axionic shift symmetries only or whether the new
set of mass parameters is equivalent, up to a
U-duality rotation, to that of
\cite{Kaloper:1999yr} in $D \leq 8$. In Section 3 we have derived the
massive S-duality rules
for the 231 parameters that arise in going from 7D to 6D. Thus the
massive S-duality in 6D is
implied by the massless S-duality in 7D. The other massive deformations
(463 for ${\bf h}$ and 24 for ${\bf iia}$) 
seem to break the S-duality in six dimensions.

It is instructive to compare the D=7 to D=6 reduction
with U-duality group $O(3,19)$ 
 with the  D=10 to D=9
reduction of IIB supergravity with U-duality group $SL(2,R)$.
In the first case we have 54 axionic shift symmetries while in the latter
case there is a single shift symmetry. In the D=10 to D=9 reduction
the SS reduction using the axionic shift symmetry relates a
domain wall in D=9 to a D7-brane in D=10. Similarly, we find that 
in the D=7 to D=6 SS reduction  each of the 54 axionic shift symmetries
relate a domain wall solution in D=6, see \eqref{h4-brane},
to a 4-brane solution in D=7.
The other SS reductions do not seem to relate in an obvious way 
brane solutions in D=6(9) with brane solutions in D=7(10). This issue 
is under present investigation \cite{new}.

The $f_{ABC}$'s  are the structure constants of a subgroup 
$G\subset O(4,20)$
in some non-standard basis \cite{Bergshoeff:1997mg,
Alonso-Alberca:2000gh}.
The $f_{ABC}$-deformations
can also be obtained directly in 6D by gauging this subgroup 
\cite{Kaloper:1999yr}. We expect the same to be true
for our general scheme. Since there are only 24 vectors
available in the supergravity multiplet
we expect that rank(G)$\leq 24$, 
though this is not manifest in the non-standard basis.
It is of interest to investigate under which conditions
a general massive deformation
can be viewed as the result of a certain gauging in the supergravity
theory and if so, which subgroup of the U-duality group is gauged.

Finally, it would be interesting to extend our work and
consider compactifications to three dimensions.
Massive supergravities in D=3 with neutral scalars were
given in \cite{Townsend}. More recently, massive supergravities in 
D=3 with charged scalars have been constructed in \cite{Per,Nicolai}.
We expect that also in 
this case there exist different classes of massive supergravities
with the mass parameters being in different representations of the U-duality
group.

\section*{Acknowledgements}

D.R. would like to thank the Institute for Theoretical Physics at Humboldt
University Berlin and the Department of Theoretical Physics at Uppsala
University for its hospitality while part of the research was being
performed. He also would like to acknowledge Michael Haack for interesting
and useful discussions, in particular concerning Section 4. P.S. would like
to thank the Institute for Theoretical Physics at the University of Groningen
for its hospitality. This work was supported by the European Commission
RTN program HPRN-CT-2000-00131, in which E.B.~and D.R.~are
associated with Utrecht University.

\appendix
\section{Notational Conventions}

Our notational conventions are similar to~\cite{Behrndt:1996si}:
space-time indices are omitted where possible but should be taken fully
antisymmetric and without factor. Thus $H = d B - V_a d V^a$
implies $H_{\mu\nu\rho} = \partial_{[\mu} B_{\nu\rho]} - V_{a \; [\mu}
\partial_\nu V^a_{\rho]}$ and
\begin{align}
  |H|^2 \equiv
    H_{\nohat{\mu} \nohat{\nu} \nohat{\rho}}
    H^{\nohat{\mu} \nohat{\nu} \nohat{\rho}},
  \qquad |\nohat{F}^a|^{2} \equiv
    \nohat{F}^a_{\nohat{\mu} \nohat{\nu}} \nohat{M}^{-1}_{ab}
    \nohat{F}^{b \; \nohat{\mu} \nohat{\nu}} \,.
\end{align}
The derivative $d$ acts from the left.
The Hodge star operator of a p-form in $D$ dimensions is defined by
\begin{align}
  (\star \nohat{H})^{\nohat{\mu_1} \cdots \nohat{\mu_p}}
  \equiv \frac{1}{p! \sqrt{-\nohat{g}}}
  \nohat{\epsilon}^{\nohat{\mu_1} \cdots \nohat{\mu_D}}
  \nohat{H}_{\nohat{\mu_{D-p+1}} \cdots \nohat{\mu_D}} \,.
\end{align}
The different dimensional Levi-Civita symbols are related by
$\nohat{\epsilon}^{\mu_1 \cdots \mu_{D-1} \underline{x}} \equiv
\epsilon^{\mu_1 \cdots \mu_{D-1}}$ when reducing over the coordinate $x$.

\section{Generalised SS Reduction on a Torus}

In this Appendix we will focus
on the reduction of the heterotic theory. The compactification of the
Heterotic string with background fluxes was originally considered in
\cite{Kaloper:1999yr}. More recently, such compactifications
have been studied in \cite{Louis:2001uy}.
Here we will rederive and extend the results
of \cite{Kaloper:1999yr}. The results of
this Appendix are needed
to construct the massive S-duality rules in Section 3.

In general, supergravity theories acquire a U-duality symmetry group upon
reduction.
This U-duality group can be used to generate masses in further
dimensionally reduced theories.
For instance, consider  the reduction on a torus, which is the product of
circles.
One can perform the compactification step-by-step, meaning circle-by-circle.
At every step there is a U-duality group that can be used to give masses
to the theory reduced further on the torus.
There are restrictions however, that impose conditions on the mass parameters.
Here we will describe this procedure for the heterotic theory but it
can be applied to any theory with a U-duality symmetry group upon reduction.

Let us start with some general remarks on the U-duality symmetry groups that
are considered in this paper.
They are all of the form $O(d,d+16)$ for some integer $d$.
The $O(d,d+16)$ index ${}_A$ can be expressed in terms of the $O(d-1,d+15)$
index ${}_a$ as $\{ {}_\bullet, {}^\bullet, {}_a \}$, while ${}^A =
\{ {}^\bullet, {}_\bullet, {}^a \}$.
This iterative procedure stops at $O(0,16)$, which has an index ${}^k$
with $k=1..16$.
All $O(d,d+16)$ indices are raised or lowered with the metric
\begin{equation}
  L_{AB} = L^{AB} = \left( \begin{array}{ccc}
  0 & 1 & 0\\
  1 & 0 & 0\\
  0 & 0 & L_{ab} \end{array} \right)\, ,
\end{equation}
with $L_{ab}$ the $O(d-1,d+15)$ metric.
The $O(0,16)$ metric is simply $L_{kl} = (- \mathbb{I}_{16})_{kl}$.
The vectors of the theory form a fundamental representation $V^A$ of
the U-duality group.
The scalars form a matrix $M^{AB}$ iteratively defined by
\begin{equation}
  M^{AB} =
  \left(
  \begin{array}{ccc}
  -e^{-2\sigma} + \ell_a \ell_b { { M}}^{ab}
  -\frac{1}{4}e^{2\sigma}{\ell}^4&
  \frac{1}{2}e^{2{\sigma}}{\ell}^2&
  {\ell}_a{ M}^{ab} - \frac{1}{2}e^{2{\sigma}}{\ell}^2
  {\ell}^b \cr
  \frac{1}{2}e^{2{\sigma}}{\ell}^2&
  -e^{2\sigma}&
  e^{2{\sigma}}{\ell}^b\cr
  { M}^{ab}{\ell}_b - \frac{1}{2}e^{2{\sigma}}{\ell}^2
  {\ell}^a &
  e^{2{\sigma}}{\ell}^a&
  { M}^{ab} - e^{2\sigma}{\ell}^a{\ell}^b
  \end{array}
  \right)\, ,
\end{equation}
where the scalars $\ell_a$ are axionic ($\ell^2 = \ell_a \ell^a$) and
$\sigma$ is a dilatonic scalar.
$M^{AB}$ parametrises the coset $O(d,d+16)/(O(d) \times O(d+16))$,
which contains $d$ dilatons and $d(d+15)$ axions.
As any element of $O(d,d+16)$ it satisfies
$M^{-1}_{AD} = L_{AB} M^{BC} L_{CD}$.
The coset $O(0,16)/(O(0) \times O(16))$ is empty: $M^{kl} = (\mathbb{I}_{16})^{kl}$.
The U-duality symmetry acts as
\begin{align}
  V^A \rightarrow \Omega^A_{\; B} V^B \,, \qquad
  M^{AB} \rightarrow \Omega^A_{\; C} \Omega^B_{\; D} M^{CD} \,, \qquad
\label{Uduality1}
\end{align}
with $\Omega^A_{\; B} \in O(d,d+16)$.
vThis group consists of $(16+2d)(15+2d)/2$ generators, of which $d(d+15)$ shift the axions with a constant: $\ell \rightarrow \ell + m$.

Let us now specify to the Heterotic theory.
Upon reduction on a torus $T^i$, the Heterotic theory obtains a
global $O(i,i+16)$ U-duality symmetry group.
This symmetry group can be used to induce mass parameters in the
lower dimensions. This has been
done for a subset of
the $O(i,i+16)$ generators, namely those that induce the axionic shifts
\cite{Kaloper:1999yr}.

Using only these global symmetries for generalised reduction, this gives
 $0, 16,16+34,16+34+54,\ldots$ mass parameters in $9, 8,7,6,\ldots$
dimensions, respectively.
However, there are constraints on these mass parameters.
Since the masses break (part of) the U-duality group, one no longer can
use all axionic shifts to generate masses in lower dimensions.
Only U-duality transformations that leave the mass parameters invariant
are true symmetries of the Lagrangian that can be used for SS reduction.
Thus the mass parameters have to satisfy certain product relations.
We would like to generalise the result of \cite{Kaloper:1999yr}
to the full U-duality group, i.e.~use
all possible generators to induce masses for the Heterotic theory
compactified on a torus.
This can be done most conventiently step-by-step, i.e.~by splitting the
torus in a product of circles and using the U-duality group at every step
\cite{Lavrinenko:1998qa}.

Let us first quote the general result.
Then we will iteratively prove that this indeed can be obtained from
a generalised SS reduction.
The massive Heterotic action in (10-i) ($i=1,2,3,4$) dimensions reads
\begin{align}
  \mathcal{L} = \tfrac{1}{2} \sqrt{-g} e^{-2\phi}
v    \big[
  & -R + 4 |d \phi|^2 -\tfrac{3}{4} |H|^2
    +\tfrac{1}{8} \text{Tr} (\mathcal{D} M \mathcal{D} M^{-1}) + \notag \\
  & - |{F}^A|^{2}
    - \tfrac{1}{12} f_{ABC} f_{DEF} M^{AD} (M^{BE} M^{CF} - 3 L^{BE} L^{CF})
\big]\, ,
\label{actionhet}
\end{align}
with the field strengths
\begin{align}
  H & = d B - V_A d V^A - \tfrac{1}{6} f_{ABC} V^A V^B V^C \,,
\notag \\
  F^A & = d V^A - \tfrac{1}{2} f_{BC}{}^A V^B V^C \,, \notag \\
  \mathcal{D} M^{AB} & = d M^{AB} - f_{CD}{}^A V^C M^{BD} -
f_{CD}{}^B V^C M^{AD} \,,
\label{fieldstrengths}
\end{align}
where $f_{ABC}$ ($A$ is the $O(i,i+16)$ index) are
fully antisymmetric expressions that
contain the mass parameters (see below). Only in special cases of
gauged supergravities can these constants be associated with
the structure constants of a Lie algebra.
The vectors $V^A$ come from three different sources: $V^h$ ($h=1..i$)
are the
Kaluza-Klein vectors, $V_h$ stem from the two-form while the $V^k$
($k=1,\cdots ,16$) are the
Yang-Mills vectors of the Heterotic theory.
The action (\ref{actionhet}) has a global $O(i,i+16)$ U-duality
pseudo-symmetry
\begin{align}
  V^A \rightarrow \Omega^A_{\; B} V^B \,, \qquad
  M^{AB} \rightarrow \Omega^A_{\; C} \Omega^B_{\; D} M^{CD} \,, \qquad
  f_{ABC} \rightarrow \Omega^A_{\; D} \Omega^B_{\; E} \Omega^C_{\; F} f_{DEF} \,,
\label{Uduality2}
\end{align}
with $\Omega^A_{\; B} \in O(i,i+16)$.
Only in the massless case, i.e. $f_{ABC} = 0$,  it is a true symmetry
acting only on fields.
This subtlety will play an important role since only true symmetries
can be used for a generalised SS reduction.

To explain the iterative derivation of (\ref{actionhet})
we start from the massless ten-dimensional heterotic action,
i.e.~$i=0$ and $f_{abc}=0$ (the gauge group will be considered to be
broken to $U(1)^{16}$).
Conventional Kaluza-Klein reduction (with no dependence on the internal
coordinates) gives the massless version of the reduced action,
i.e.~the action (\ref{actionhet}) with
$f_{AB}{}^C$ set equal to zero.
We now will use the full U-duality symmetry to generate masses in lower
dimensions.
To see which structure coefficients can be made non-zero by generalised
reduction we will iteratively perform the compactification step-by-step.
Thus the action \eqref{actionhet} is compactified on a circle, using the
most general reductions relations consistent with the U-duality group.
We employ the following reduction relations expressing the
(10-j)-dimensional ($j = i-1= 0,1,2,3$) fields in terms of the
(9-j)-dimensional
fields:
\begin{align}
  \begin{array}{rcl}
  \underline{\text{(10-j)D}} && \underline{\text{(9-j)D}} \\
  {\nohat g}_{\underline{zz}} &=& - e^{-4\sigma/{\sqrt 3}}\, ,\\
  {\nohat g}_{{\underline z}\mu} &=& -e^{-4\sigma/{\sqrt 3}}V^\bullet_\mu\, ,\\
  {\nohat g}_{\mu\nu} &=& g_{\mu\nu} - e^{-4\sigma/{\sqrt 3}}V^\bullet_\mu
V^\bullet_\nu\, ,\\
  {\nohat B}_{\mu\nu} &=& B_{\mu\nu} + V^\bullet_{[\mu}(V_\bullet)_{\nu]} + \ell^a
V_{[\mu}^bV^\bullet_{\nu]}L_{ab}\, ,\\
  {\nohat B}_{\underline z \mu} &=& (V_\bullet)_\mu -\tfrac{1}{2}\ell^a
V_\mu^b L_{ab}\, ,\\
  \nohat \phi &=& \phi - \tfrac{1}{{\sqrt 3}} \sigma\, ,\\
  {\nohat V}_\mu^a &=& \Omega^{a}_{\; b} (z) (V_\mu^b +\ell^b V^\bullet_\mu) \, ,\\
  {\nohat V}_{\underline z}^a &=& \Omega^{a}_{\; b} (z) \ell^b \, ,\\
  {\nohat M}^{ab} &=& \Omega^{a}_{\; c} (z) \Omega^{b}_{\; d} (z) M^{cd}\, .
  \end{array}
\label{red76}
\end{align}
with $\Omega^{a}_{\; b} (z) \in O(j,j+16)$ ($a = 1..2j+16)$
being the only $z$-dependence
(the coordinate of the circle) on the right-hand side.
These are the usual circle reduction formulae apart from the
'reduction transformation' $\Omega^{a}_{\; b} (z)$.
The single-valuedness of the reduced theory (independence of $z$)
follows from the U-duality symmetry \eqref{Uduality2} provided
\cite{Lavrinenko:1998qa}
\begin{align}
  \Omega^{ab} \partial_z \Omega_{ac} (z)
  = - (m^{(j)})^{b}_{\; c} \,,
\end{align}
with $(m^{(j)})_{ab}$ an arbitrary antisymmetric (16+2j)x(16+2j) matrix.
{From} this it follows that
\begin{align}
  \Omega^{a}_{\; b} (z) = e^{- (m^{(j)})^{a}_{\; b} z} \,.
\label{omegaz}
\end{align}
Reducing with these generalised relations we find the following field
strengths for the vector fields:
\begin{align}
  F^a & = d V^a - (m^{(j)})^a_{\; b} V^\bullet V^b - \tfrac{1}{2}
f_{bc}{}^c V^b V^c \,, \notag \\
  F^{\bullet} & = d V^\bullet \,, \notag \\
  F_{\bullet} & = d V_\bullet - \tfrac{1}{2} (m^{(j)})_{ab} V^a V^b \,,
\end{align}
where $V^\bullet$ is the Kaluza-Klein vector of the circle and $V_\bullet$
comes
from the two-form $B$.
These can be combined into $F^A$ with ${}^A$ the $O(i,i+16)$ index
$\{ {}^\bullet, {}_\bullet, {}^a \}$ where $i=j+1$.
Thus we find the (10-i) dimensional non-Abelian field strengths
\eqref{fieldstrengths} with non-zero structure coefficients
\begin{align}
  f_{ABC}: \qquad f_{abc} \,, \qquad f_{ab\,\bullet} = (m^{(j)})_{ab} \,.
\end{align}
The scalar sector is somewhat more complicated but also reproduces
\eqref{fieldstrengths} with the above structure coefficients.

There are restrictions to this generalised SS reduction scheme.
If in the higher-dimensional theory mass parameters are present, the U-duality symmetry group is (partly) broken.
The only true symmetries are those that leave the structure coefficients invariant, i.e.
\begin{align}
  f_{abc} = \Omega_a^{\; d} \Omega_b^{\; e} \Omega_c^{\; f} f_{def} \,.
\label{unbrokenUduality}
\end{align}
Thus not all U-duality transformations can be used to generate masses.
Plugging in the 'reduction transformation' \eqref{omegaz} one obtains relations between the mass parameters.
These exactly boil down to the Jacobi identity for the structure coefficients of the reduced theory \cite{Kaloper:1999yr}:
\begin{align}
  f_{A[B}{}^C f_{DE]C} = 0 \,.
\label{Jacobi}
\end{align}

In summary, there exists an iterative step-by-step prescription to obtain
families of mass parameters in the Heterotic theory compactified on  a torus
$T^i$.
At the j-th step ($j=0,1,2,i-1)$ one can introduce a
family $m^{(j)}$ with
$(16+2j)(15+2j)/2$ independent mass parameters.
These can be grouped into structure coefficients of the lower-dimensional
theory:
\begin{align}
  f_{ABC}: \qquad f_{abj} = (m^{(j)})_{ab} \,,
\end{align}
where ${}_A = \{{}_j, {}^j, {}_a \}$ is the index of the U-duality
group of the reduced theory.
When introducing more than one family of mass parameters one has restrictions
\begin{align}
  f_{A[B}{}^C f_{DE]C} = 0 \,,
\end{align}
which follow from the fact that mass parameters break (part of) the U-duality symmetry group.
For only one family of non-zero masses these are identically satisfied.
Of course it is possible to reshuffle the mass parameters by field redefinitions.
In this way one can redefine the structure coefficients by
\begin{align}
  f_{ABC} \rightarrow \Omega_A^{\; D} \Omega_B^{\; E} \Omega_C^{\; F}
f_{DEF} \,,
\end{align}
with $\Omega_A^{\; B}$ an element of the U-duality group.
Thus there is the issue of the number of independent parameters.
Finally, our reduction scheme is more general than that of
\cite{Kaloper:1999yr}. For instance, using only the axionic shift symmetries
the Heterotic theory cannot induce
mass parameters in nine dimensions while using the full U-duality, as
we do here, one can generate massive deformations already in nine dimensions.


\end{document}